\documentclass[10pt,letterpaper]{article}
\usepackage{opex3}
\usepackage{upgreek}

\begin{document}

\title{Bright filter-free source\\of indistinguishable photon pairs}

\author{F.~Wolfgramm,$^{1}$ X.~Xing,$^2$ A.~Cer\`{e},$^1$ A.~Predojevi\'{c},$^1$ A.~M.~Steinberg,$^2$ and M.~W.~Mitchell$^1$}
\address{$^1$ICFO - Institut de Ciencies Fotoniques, Mediterranean
Technology Park,
\\
08860 Castelldefels (Barcelona), Spain
\\
$^2$Centre for Quantum Information \& Quantum Control and
Institute for Optical Sciences,
\\
Dept. of Physics, 60 St. George St., University of Toronto, Toronto,
ON, Canada, M5S 1A7}
\email{florian.wolfgramm@icfo.es}



\begin{abstract}
We demonstrate a high-brightness source of pairs of
indistinguishable photons based on a type-II phase-matched
doubly-resonant optical parametric oscillator operated far below
threshold. The cavity-enhanced down-conversion output of a PPKTP
crystal is coupled into two single-mode fibers with a mode coupling
efficiency of 58\%. The high degree of indistinguishability between
the photons of a pair is demonstrated by a Hong-Ou-Mandel
interference visibility of higher than 90\% without any filtering at
an instantaneous coincidence rate of 450~000 pairs/s per mW of pump
power per nm of down-conversion bandwidth. For the degenerate
spectral mode with a linewidth of 7~MHz at 795~nm a rate of 70
pairs/(s~mW~MHz) is estimated, increasing the spectral brightness
for indistinguishable photons by two orders of magnitude compared to
similar previous sources.
\end{abstract}


\section{Introduction}

Indistinguishable photons exhibit non-classical interference, the
basis for applications including entanglement generation
\cite{Ou1988}, linear-optics quantum computing \cite{Knill2001}, and
super-resolving phase measurements \cite{Mitchell2004}.  Many
down-conversion sources \cite{Kwiat1995, Fedrizzi2007, Kuzucu2008}
achieve indistinguishability by spectral and/or spatial filtering,
at the cost of reduced efficiency. In contrast, cavity-enhanced
down-conversion \cite{Ou1999, Lu2003, Wang2004, Kuklewicz2006,
Neergaard-Nielsen2007, Scholz2007} promises to produce photon pairs
with cavity-defined, and thus controllable, spatial and spectral
characteristics. In this spirit, Ou \emph{et al.} \cite{Ou1999} used
a type-I optical parametric oscillator (OPO) far below threshold to
produce a two-photon state that was later shown to be mode-locked
and consisting of indistinguishable photons by observing the
Hong-Ou-Mandel effect with moderate visibility
\cite{Lu2003,Hong1987}. A high-purity source of heralded single
photons was recently demonstrated by Neergaard-Nielsen \emph{et al.}
\cite{Neergaard-Nielsen2007} with a sub-threshold type-I OPO and
filters to select heralding photons from a single mode. Kuklewicz
\emph{et al.} \cite{Kuklewicz2006} used a type-II OPO below
threshold with temporal filtering (selection of nearly-coincident
pairs), to demonstrate a high-brightness source with up to 76.8\%
Hong-Ou-Mandel interference visibility. The source that we
demonstrate here avoids filtering altogether: all output of the
source is used, giving a brightness tens to hundreds of times higher
than earlier sources. At the same time, high-quality pairs are
produced, as demonstrated by a high-visibility Hong-Ou-Mandel dip.
The source is also tunable to a rubidium resonance, allowing studies
of light-atom interactions with photon pairs.

Despite energy conservation and phase-matching requirements,
parametric down-conversion is generally under-constrained, allowing
emission into a range of energy and momentum states.  Emitted
photons are correlated in energy and momentum, providing both an
entanglement resource and a challenge to efficient collection.  By
their nature, position-momentum entangled states cannot be
efficiently collected into single-mode optics.  Correlations between
observed and unobserved variables can also render the photons
distinguishable and destroy quantum interference \cite{Adamson2007}.
While filters can "erase" the distinguishing information, they
change the nature of the source, from one which produces entangled
pairs (and nothing else) to one which sometimes produces unpaired
photons. Also, the heralding efficiency is reduced, exponentially
decreasing the success probability in a multi-photon heralded
experiment.  Rather than use filters, we place the down-conversion
within a resonant cavity \cite{Ou1999, Kuklewicz2006}. As with other
spontaneous processes \cite{Heinzen1987, Martini1987, Kuhn2002},
this constrains the emission to the cavity modes and enhances the
emission rate into those modes. The cavity is designed and
stabilized for simultaneous resonance on all longitudinal modes
within the phase-matching bandwidth, producing both a very bright
source and pair indistinguishability.

\section{Experimental setup}
\begin{figure}[t]
\centering
\includegraphics[width=8.3cm]{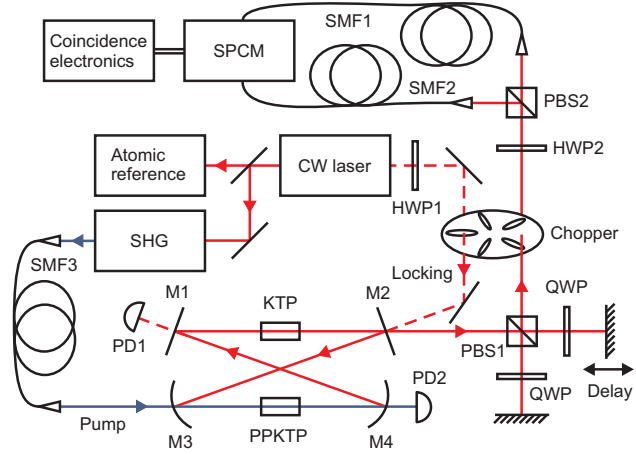}
\caption{Experimental Setup. PPKTP, phase-matched nonlinear crystal;
KTP, compensating crystal; M1-4, cavity mirrors; PBS, polarizing
beam splitter; HWP, half wave plate; QWP, quarter wave plate; SMF,
single-mode fiber; PD, photodiode\label{img:Setup}}
\end{figure}

As laser source we use a frequency-doubled diode laser system
(Toptica TA-SHG 110). The laser wavelength is stabilized to the
D$_1$ transition of atomic rubidium at 795~nm and then frequency
doubled to generate the 397.5~nm pump that is passed through a
mode-cleaning single-mode fiber and is focused into the center of a
20 mm-long periodically-poled KTiOPO$_4$ \mbox{(PPKTP)} crystal in a
cavity, forming the OPO (Fig.\ \ref{img:Setup}). A pump beam waist
of 30~$\upmu$m is achieved with a telescope. This beam waist was
chosen to be larger than the optimum for degenerate down-conversion
according to Boyd and Kleinman \cite{Boyd1968} in order to reduce
possible effects of thermal lensing \cite{LeTargat2005} and
gray-tracking \cite{Boulanger1994}. The crystal is poled for type-II
degenerate down-conversion, and produces orthogonally-polarized
signal and idler photons.  Due to crystal birefringence, these
photons experience temporal walk-off which would, if un-compensated,
render the photons temporally distinguishable.  A second KTP crystal
of the same length and crystal cut, but not phase-matched and
rotated about the beam direction by 90$^{\circ}$, is added to the
long arm of the cavity in order to introduce a second walk-off equal
in magnitude but opposite in sign \cite{Kuklewicz2006}.

The ring cavity is formed by two flat mirrors (M1, M2) and two
concave mirrors (M3, M4) with a radius of curvature of 100~mm. The
effective cavity length of 610~mm corresponds to a free spectral
range (FSR) of 490~MHz. This geometry provides a beam waist of
42~$\upmu$m for the resonant down-converted beam at the center of
the crystal, which matches the 30~$\upmu$m pump beam waist. Cavity length is controlled by a piezoelectric
transducer on mirror M1. The output coupler M2 has a reflectivity of
93\% at 795~nm. All other cavity mirrors are highly reflecting
(R~$>$~99.9\%) at 795~nm and highly transmitting at 397.5~nm
(R~$<$~3\%) resulting in a single-pass through the nonlinear crystal
for the blue pump beam. The crystal endfaces are AR coated for
397.5~nm and 795~nm. The measured cavity finesse of 70 results in a
cavity linewidth of 7~MHz.

While the walk-off per round trip is compensated by the KTP crystal,
there is an uncompensated walk-off of in average half a
crystal-length, because of the different positions inside the PPKTP,
where a photon pair could be generated. This leads to a remaining
temporal distinguishability at the output of the cavity that is
completely removed by delaying the horizontally polarized photon of
each pair with a Michelson-geometry compensator: a polarizing beam
splitter, retro-reflecting mirrors, and quarter wave-plates set to
45$^{\circ}$ introduce an adjustable delay while preserving spatial
mode overlap. After recombination the pairs are sent through a half
wave plate (HWP2) that together with PBS2 determines the measurement
basis. Both output ports of PBS2 are coupled into single-mode fibers
(SMF) connected to single photon counting modules (Perkin Elmer
SPCM-AQ4C). The pulse events are registered and processed by
coincidence electronics (FAST ComTec P7888) with a resolution of
1~ns.

The OPO cavity is actively stabilized by injecting an auxiliary
beam, derived from the diode laser, into the cavity via the output
coupler (M2). This light is detected in transmission by a photodiode
(PD1). Frequency modulation at 20~MHz is used to lock to the peak of
the cavity transmission. To eliminate the background noise caused by
this auxiliary beam and to protect the SPCMs, the locking and
measuring intervals are alternated using a mechanical chopper at a
frequency of about 80~Hz with a duty cycle of 24\%.

To achieve degeneracy of the H/V modes, we set the polarization of
the auxiliary beam to $45^{\circ}$ and measure the cavity
transmission for H- and V-polarized components. The transmission
peaks for the two polarizations are overlapped by temperature tuning
of the compensating KTP crystal, whereas the temperature of the
PPKTP crystal is kept stable at the optimal phase-matching
temperature for degenerate operation at $42.0^{\circ}$C. Both
crystals are temperature controlled with a long-term stability of
better than 5~mK corresponding to an overlap between the
transmission spectra for H- and V-polarized components of better
than 95\%.

The pump power after the cavity is measured by collecting a constant
fraction on a photodiode (PD2). This system is calibrated against
the total power passing through the PPKTP crystal as measured by a
power meter (Coherent FieldMate). Over the duration of any given
measurement, the power was stable to within 5\%. We used a typical
optical pump power of 200~$\upmu$W to reduce the probability of
generation of two pairs within the cavity ring-down time. The
optimization of the pump beam mode-matching was performed by
maximizing the count rates on the single-photon detectors.

\section{Arrival-time correlation measurement}

Without correcting for any losses, the photon rate in each arm
$(R_{\rm SMF1, SMF2})$ was measured to be 142~000 counts/s during
the measurement period (when the chopper was open) with a
coincidence rate of 34~000 pairs/s. The unavoidable accidental
coincidence rate $R_{\rm acc}$ in the coincidence time window of
$\tau=256~$ns is calculated by $R_{\rm acc}=R_{\rm SMF1}~R_{\rm
SMF2}~\tau=5~000$ pairs/s, resulting in a corrected coincidence rate
of 29~000 pairs/s, that is, a collection efficiency of~20\%.
\\
The FWHM of the crystal phase-matching bandwidth is calculated by
$1/(|k_s'-k_i'|L)=148~$GHz, with the crystal length $L$ and the
$k$-vectors for signal and idler photons \cite{Fedrizzi2008}.
Considering this bandwidth a spectral brightness of 450~000
pairs/(s~mW~nm) \cite{Footnote1} is calculated, brighter by a factor
of 1.6 compared to the single-pass SPDC source in
\cite{Fedrizzi2007} with a coincidence rate of 273~000
pairs/(s~mW~nm).
\begin{figure}[t]
\centering
\includegraphics[width=9.3cm]{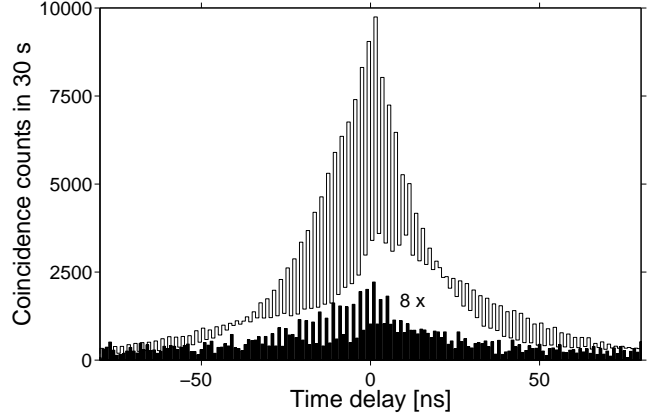}
\caption{Histogram of difference between signal and idler arrival
times in the $\pm45^{\circ}$ basis (multiplied by 8, lower curve)
and in the H/V basis (upper curve) without path difference between signal and idler beam paths\label{img:Histogram}}
\end{figure}

Within the crystal bandwidth the output spectrum consists of roughly
600~modes, the degenerate one at the rubidium D$_1$ line. The modes
are spaced by a FSR of 490~MHz, each of them having a bandwidth of
7~MHz. For the degenerate mode, which contains about 1/300 of the
total output, the photon coincidence rate is estimated to be 70
pairs/(s~mW~MHz), two orders of magnitude higher than the rate of
0.7 pairs/(s~mW~MHz) reported in \cite{Kuklewicz2006}.

Taking into account the limited quantum efficiency of the detectors
(49\%), the single-mode fiber coupling efficiency (58\%), the escape
efficiency of the cavity (82\%) and the overall transmission through
all optical elements after the cavity (90\%) a conditional detection
efficiency of 21\% is expected. The fact that the measured
efficiency agrees with this expectation proves that the emission
from the cavity itself is in a single spatial mode. Given this parameter, we
estimate the true pair production rate to be $3.4\times 10^{6}$
pairs/(s~mW).

We measured the arrival time difference between detection events on
the two SPCMs in a time window of 256~ns over 30 seconds in the H/V
and $\pm 45^{\circ}$ bases (Fig.\ \ref{img:Histogram}). The time
correlation graph shows the typical double-exponential decay
reflecting the ring-down time of the cavity. The time delay between
the photons of a pair is always an integer multiple of the round
trip time, resulting in the comb structure of Fig.\
\ref{img:Histogram}, in which alternating bins have low and high
count rates. The contrast of this oscillation is modulated due to
the sampling resolution of 1.0~ns and the 2.03~ns round trip time,
vanishing every 31 peaks as can be seen in Fig.\
\ref{img:Histogram}. These results agree with the theory given in
\cite{Kuklewicz2006} for the case of compensated birefringence.

\section{Hong-Ou-Mandel measurement}

When the relative delay between the two photons of a pair is
changed, their degree of temporal indistinguishability is varied and
the Hong-Ou-Mandel effect \cite{Hong1987} can be observed. When we
measure in the $\pm45^{\circ}$ basis and with no delay, signal and
idler photons of a pair impinging on PBS2 are indistinguishable and
exit on the same output port of PBS2 leading to a drop in the
coincidence rate as shown in Fig.\ \ref{img:Histogram}.
\begin{figure}[b]
\centering
\includegraphics[width=9.3cm]{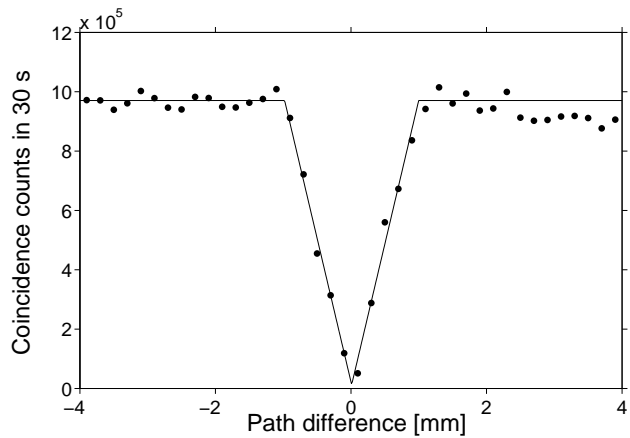}
\caption{Hong-Ou-Mandel Dip. Experimental data and triangular fit
function.\label{img:HOM}}
\end{figure}

The coincidence rate in the $\pm45^{\circ}$ basis was measured for
different mirror positions over a range of 8~mm with a step width of
0.2~mm, accumulating coincidence counts at each point for 30
seconds. All coincidences within the time window of $\tau=256~$ns
are counted and the results are shown in Fig.\ \ref{img:HOM}.
Accidentals due to double pairs are subtracted and the data are
corrected for power fluctuations in the pump. The statistical error
bars are too small to be displayed. As expected for an unfiltered
type-II SPDC source, the HOM dip shows a triangular shape
\cite{Grice1997}. The model that we use for the fit is based on the
theory given in \cite{Fedrizzi2008}. The coincidence rate $R_{\rm
coin}$ is given in terms of the difference $\Delta l$ between signal
and idler paths, the average coincidence rate for large time
differences $R_{\rm avg}$ and a parameter $\zeta$ with $\zeta=4/(L
|k_s'-k_i'|)$, that depends on the crystal length $L$ and the
$k$-vectors for signal and idler photons:
\begin{equation}
R_{\rm coin}(\Delta l)=R_{\rm avg} \left(1-\wedge \left(\frac{\Delta
l \zeta}{2c} \right) \right)
\end{equation}
The function $\wedge(x)$ takes the value $\wedge(x)=1-|x|$ for
$|x|<1$ and $\wedge(x)=0$ elsewhere. The theoretical prediction of
the base-to-base width of the triangle $4c/\zeta=2.03$ mm agrees
well with the fitted value of 2.0 mm.
\\
The drop of the coincidence rate for path differences larger than
+2.5~mm is due to a change in coupling efficiency to the single-mode
fibers, as the efficiency was optimized for translation stage
positions close to the bottom of the dip. Therefore, data points
over +2.5~mm were disregarded for the fit. The fit function displays
a visibility defined by $V=(C_{\rm max}-C_{\rm min})/(C_{\rm
max}+C_{\rm min})$ of 96\% with subtraction of accidental counts and
83\% without; the lowest point measured directly shows a visibility
of 90\%. To reduce the rate of accidental counts due to double pair
generation even more, the HOM dip was also measured for a very low
pump power of 12~$\upmu$W. For this measurement the visibility for
the lowest data point is 95\% with subtraction of the accidentals
and 90\% for the raw data. This visibility clearly indicates the
non-classical character of the down-converted photon pairs and their
indistinguishability \cite{Ou1989}. It should be noted that all our
measurements were done without any spectral filtering.

\section{Conclusion}
In conclusion, we have demonstrated a high-brightness, highly
efficient source of pairs of indistinguishable photons using a
type-II OPO. Compared to other schemes it shows a higher degree of
indistinguishability and a higher flux. Our setup achieves
indistinguishability without any spectral or spatial filtering,
which allows for the first time efficient coupling of high-quality
photon pairs into single-mode fibers. A Hong-Ou-Mandel dip was
measured with a visibility of over 90\%. The source can easily be
extended to provide beams of polarization-entangled photons by using
an ordinary 50-50 beam splitter. The degenerate spectral mode with a
linewidth of 7~MHz at 795~nm is estimated to contain 70
pairs/(s~mW~MHz), increasing the spectral brightness for pairs of
indistinguishable photons by two orders of magnitude over the, to
our knowledge, spectrally brightest comparable source reported so
far. Depending on the application, filter cavities after the OPO
could be used to isolate this degenerate mode or the unfiltered
output could be directly applied on a frequency-selective system
such as atoms. Since the spectral properties of this mode match
perfectly the natural linewidth of the D$_1$ transition of atomic
rubidium, the presented source provides photon pairs for efficient
interaction with atomic systems. This is an essential requirement
for the realization of an interface between light and matter on the
single-photon level for quantum memories and quantum repeaters.

\section*{Acknowledgments}
This work was supported by an ICFO-OCE Collaborative Research
Program, by the Spanish Ministry of Science and Education under the
FLUCMEM project (Ref. FIS2005-03394) and the Consolider-Ingenio 2010
Project ``QOIT'', by NSERC, the Canadian Institute for Photonic
Innovations, Ontario Centres of Excellence and QuantumWorks. F.~W.
and A.~P. are supported by the Commission for Universities and
Research of the Department of Innovation, Universities and
Enterprises of the Catalan Government and the European Social Fund.


\begin{thebibliography}{99}
\bibitem{Footnote1} Although we express the efficiency as pairs/(s mW nm) or pairs/(s mW MHz) for comparison with previous sources and in preparation for future experiments with atoms, in this experiment the pump power is not a limiting factor.
\end{thebibliography}
\end{document}